\documentclass[11pt]{article}

\usepackage{amssymb,latexsym,amsmath}

\newtheorem{thm}{Theorem}[section]
\newtheorem{cor}[thm]{Corollary}
\newtheorem{lem}[thm]{Lemma}

\begin{document}

\def\prf{\noindent{\bf Proof.}\ }
\def\prfe{\hspace*{\fill} $\Box$

\smallskip \noindent}

\title{Stability of disk-like galaxies---Part I: Stability via reduction}
\author{Roman Fi\v{r}t, Gerhard Rein\\
         Mathematisches Institut der
         Universit\"at Bayreuth\\
         D 95440 Bayreuth, Germany}
\maketitle

\begin{abstract}
We prove the existence and stability of flat steady states of the
Vlasov-Poisson system, which in astrophysics are used as models
of disk-like galaxies. We follow the variational approach
developed by {\sc Guo} and {\sc Rein} \cite{GR1,GR2,GR4} for this type of
problems and extend previous results of {\sc Rein} \cite{reinflat}.
In particular, we employ a reduction procedure which relates the stability
problem for the Vlasov-Poisson system to the analogous question
for the Euler-Poisson system. 
\end{abstract}

\section{Introduction}
In astrophysics, galaxies or globular clusters are often modeled as
a large ensemble of particles (stars) interacting only by the gravitational 
field which they create collectively. 
In such systems collisions among particles
are sufficiently rare to be neglected. Hence 
in a nonrelativistic setting the particles move on trajectories determined 
by Newton's equations of motion
\[
\dot X=V,\
\dot V=-\nabla_{X}U(t,X),
\]
where $U(t,X)$ denotes the gravitational potential of the ensemble, 
$t\in\mathbb{R}$ is time, and $X,V\in\mathbb{R}^{3}$ denote position 
and velocity; it is assumed that all particles have the same mass
which is normalized to unity. To describe the time evolution of the 
ensemble, the density function 
$F:\mathbb{R}\times\mathbb{R}^{3}\times\mathbb{R}^{3}\to\mathbb{R}^{+}_{0}$
on phase space is used. 
It obeys a conservation law known as the Vlasov equation,
\[
\frac{\partial F}{\partial t}+V\cdot\nabla_{X}F
-\nabla_{X}U \cdot\nabla_{V}F=0,
\]
which can be understood as an incompressibility condition of the
"fluid" in phase space, also known as Liouville's
theorem. The gravitational potential $U$ is induced by the spatial density
\[
R (t,X):=\int_{\mathbb{R}^{3}} F(t,X,V)\,\mathrm{d}V
\]
via Newton's law of gravity
\[
U (t,X) = -\int_{\mathbb{R}^3}\frac{R (t,Y)}{|X-Y|}\,\mathrm{d}Y.
\]
In the present paper we are interested in a situation
where extremely flattened objects such flat galaxies are to be modeled.
We therefore assume that all
particles are concentrated in a plane, say the $(x_{1},x_{2})$-plane,
with velocity vectors tangent to it. If this holds initially and
if the only force acting on the particles is their mutual gravitational
attraction, then the particles stay in that plane, and
we can introduce a new, flat particle density
$f:\mathbb{R}\times\mathbb{R}^{2}\times\mathbb{R}^{2}\to\mathbb{R}_{0}^{+}$,
which is related to the density on the full phase space through
\[
F(t,X,V)=f(t,x,v)\delta(x_{3})\delta(v_{3}).
\]
Here $\delta$ denotes the Dirac distribution, and $X=(x,x_{3}),\;
V=(v,v_{3})$ with $x,v \in \mathbb{R}^{2}$.  
However, the particles still interact by the three
dimensional Newtonian gravitational potential. Therefore, 
the Vlasov-Poisson system for $f$ reads
\[
\frac{\partial f}{\partial t}+v\cdot\nabla_{x}f
-\nabla_{x}U \cdot\nabla_{v}f = 0,
\]
\[
U (t,x) = -\int_{\mathbb{R}^2}\frac{\rho (t,y)}{|x-y|}\,\mathrm{d}y,
\]
\[
\rho (t,x) = \int_{\mathbb{R}^2}f(t,x,v)\,\mathrm{d}v,
\]
where $x, v \in \mathbb{R}^{2}$.
We refer to this system as the flat Vlasov-Poisson system.
From the mathematics point of view it consists of the two
dimensional Vlasov equation coupled with the
$1/|x|$-type potential typical for three
space dimensions. Since the $1/|x|$-singularity is
integrated only over $\mathbb{R}^{2}$ this  kind of coupling
makes the system more singular and mathematically more difficult
to analyze than the regular, three dimensional system. 

The aim of the present investigation is to prove the existence
of a large class of non-linearly stable steady states of the
flat Vlasov-Poisson system. To do so we follow the approach
developed by {\sc Guo} and {\sc Rein}  \cite{GR1,GR2,GR4}
in the regular, three dimensional situation.
We prove that under suitable assumptions on a prescribed
function $\Phi: [0,\infty[ \to  [0,\infty[$ the energy-Casimir
functional
\begin{eqnarray*}
\mathcal{H}_{\mathcal{C}}(f) 
&=& 
\frac{1}{2}\iint |v|^{2}f(x,v)\,\mathrm{d}v\,\mathrm{d}x
-\frac{1}{2}\iint\iint\frac{f(x,v)\, f(y,w)}{|x-y|}
\,\mathrm{d}v\,\mathrm{d}x\,\mathrm{d}w\,\mathrm{d}y \\
&&
{} +
\iint \Phi(f(x,v))\,\mathrm{d}v\,\mathrm{d}x
\end{eqnarray*}
has a minimizer $f_0$ subject to the constraint 
\[
\iint f(x,v)\,\mathrm{d}v\,\mathrm{d}x = M,
\]
where $M>0$, the total mass of the resulting steady state, is prescribed.
The Euler-Lagrange relation for this variational problem implies that
\[
f_0(x,v) = \phi(E).
\]
Here the particle energy $E$ is defined as
\[
E(x,v)=\frac{1}{2}|v|^{2}+U_0(x)
\]
with $U_0$ the potential induced by $f_0$, and the function $\phi$ is
determined by $\Phi$ and a Lagrange multiplier. The point now is that
for the time-independent potential $U_0$ the particle energy $E$
and hence also $\phi(E)$ is constant along particle trajectories and
hence a solution of the time-independent Vlasov equation. Hence
$f_0$ is a steady state of the flat Vlasov-Poisson system.  
The fact that $f_0$ minimizes the energy-Casimir functional
$\mathcal{H}_{\mathcal{C}}$
can then be used to derive a non-linear stability property for 
this steady state. 

In \cite{reinflat} this approach has already been used
to construct stable steady states of the flat Vlasov-Poisson system.
In the present paper we obtain a number of improvements and extensions
of this earlier result. Firstly, we use a reduction procedure
for proving the existence of a minimizer of $\mathcal{H}_{\mathcal{C}}$.
This approach is mathematically more elegant and adequate, since
the reduced functional lives on the set of spatial densities
$\rho$, and the main difficulty in the variational problem lies
in the potential energy part which does not really depend
on $f$ but only on the spatial density induced by $f$. More importantly,
the reduced variational problem is of interest in its own right
since it provides a stability result for the flat Euler-Poisson system
which is the fluid dynamical analogue of the kinetic Vlasov-Poisson system.
For the reduction procedure to work the function $\Phi$ has to satisfy
certain growth conditions. An example of a steady state which violates this
growth condition is the so-called Kuzmin disk which is known in 
the astrophysics literature and was not covered by previous results.
The Kuzmin disk will be investigated in a companion paper \cite{F}.
Secondly, in \cite{reinflat} the perturbations admissible in the stability
result had to be supported on the plane and in addition had to be
spherically symmetric. In the present paper we remove the latter,
unphysical restriction. It is desirable to remove also the restriction
that the perturbations have to live in the plane, but that is much harder 
and is still under investigation. Lastly, we relax the assumptions
on $\Phi$ the main one being that $\Phi$ be strictly convex so that
we cover a larger class of steady states, and we obtain stability estimates
in stronger norms than were obtained previously.

The paper proceeds as follows. In the next section we introduce various
functionals and the reduced version of the variational problem,
and we establish the connection between the original and the reduced
variational problem. In the third section we establish the existence of
a minimizer to the reduced problem using a concentration-compactness
argument; notice that the variational problem---both reduced and original---is
non-trivial since the energy-Casimir functional is not convex and is defined
on functions supported on $\mathbb{R}^{2}$ or $\mathbb{R}^{4}$ respectively.
In Section~4 we derive our stability result, and in the final section
we consider the stability result for the Euler-Poisson system which
arises from the reduced functional.
   
\section{ Energy-Casimir functionals and reduction}
For $\rho=\rho(x)$ measurable we define the induced gravitational potential
and  potential energy as
\[
U_{\rho}(x)
:=
-\int\frac{\rho (y)}{|x-y|}\,\mathrm{d}y,
\]
\[
E_{\mathrm{pot}}(\rho)
:=
\frac{1}{2}\int U_{\rho}(x)\rho (x)\,\mathrm{d}x
=-\frac{1}{2}\iint\frac{\rho (x)\rho (y)}{|x-y|}\,\mathrm{d}y\,\mathrm{d}x;
\]
in the sequel integrals $\int$ without a subscript always extend
over $\mathbb{R}^{2}$.
It will also be useful to introduce the bilinear form which corresponds 
to the potential energy, i.e., 
for $\rho,\sigma:\mathbb{R}^{2} \to \mathbb{R}$ measurable,
\[
\langle \rho,\sigma \rangle_{\mathrm{pot}}:=
\frac{1}{2}\iint\frac{\rho (x)\sigma(y)}{|x-y|}\,\mathrm{d}y\,\mathrm{d}x,
\]
so that in particular 
$E_{\mathrm{pot}}(\rho) =- \langle \rho,\rho \rangle_{\mathrm{pot}}$.
For the convenience of the reader we collect the main estimates for
potentials, potential energies, and the above bilinear form, 
which we will need.
\begin{lem}
\label{epotes}
If $\rho\in L^{4/3}(\mathbb{R}^{2})$, then $U_{\rho}\in L^{4}(\mathbb{R}^{2})$, 
and there exists a constant $C>0$ such that for all 
$\rho\in L^{4/3}(\mathbb{R}^2)$ the estimates 
\[
||U_{\rho}||_{4}\le C||\rho||_{4/3},\quad
-E_{\mathrm{pot}}(\rho)\le C||\rho||_{4/3}^{2}
\]
hold. The bilinear form $\langle \cdot,\cdot \rangle_{\mathrm{pot}}$
defines a scalar product on $L^{4/3}(\mathbb{R}^{2})$
with induced norm
\[
||\rho||_{\mathrm{pot}}:=\langle\rho,\rho\rangle_{\mathrm{pot}}^{1/2}
=(-E_{\mathrm{pot}}(\rho))^{1/2},
\] 
in particular,
\[
\langle \rho,\sigma \rangle_{\mathrm{pot}}
\leq (E_{\mathrm{pot}}(\rho)\,E_{\mathrm{pot}}(\sigma))^{1/2}
=||\rho||_{\mathrm{pot}} ||\sigma||_{\mathrm{pot}}.
\]
\end{lem}
\prf
Since $1/|\cdot|\in L_{w}^{2}(\mathbb{R}^{2})$, the weak $L^2$ space, 
the assertions on $U_\rho$
follow by the generalized Young's inequality \cite[4.3]{LL}.
The estimate for the potential energy is nothing but
the Hardy-Littlewood-Sobolev inequality \cite[4.3]{LL}
and follows by H\"older's inequality, and so does 
the fact that $\langle \cdot,\cdot \rangle_{\mathrm{pot}}$ is defined
on $L^{4/3}(\mathbb{R}^{2})$.
The positive definiteness of $\langle \cdot,\cdot \rangle_{\mathrm{pot}}$
can be shown exactly like the positivity of the Coulomb energy in the
three dimensional case, cf.~\cite[9.8]{LL}.
\prfe
Let $f=f(x,v)$ be a measurable function on phase space. 
We define the induced spatial density, gravitational potential, 
and potential energy as
\[
\rho_{f}(x)
:=
\int f(x,v)\,\mathrm{d}v,\quad
U_f:=U_{\rho_f},\quad
E_{\mathrm{pot}}(f):=E_{\mathrm{pot}}(\rho_f).
\]
In addition, we define the kinetic energy
\[
E_{\mathrm{kin}}(f)
:=
\frac{1}{2}\iint |v|^{2}f(x,v)\,\mathrm{d}v\,\mathrm{d}x,
\]
the so-called Casimir functional
\[
\mathcal{C}(f):=\iint \Phi(f(x,v))\,\mathrm{d}v\,\mathrm{d}x
\]
with $\Phi:[0,\infty[ \to [0,\infty[$ prescribed,
and the energy-Casimir functional
\[
\mathcal{H}_{\mathcal{C}}(f)
:=\mathcal{C}(f)+E_{\mathrm{kin}}(f)+E_{\mathrm{pot}}(f).
\]
The total energy
$E_{\mathrm{kin}}+E_{\mathrm{pot}}$
as well as the Casimir functional $\mathcal{C}$ and hence also their sum
$\mathcal{H}_{\mathcal{C}}$ are conserved along sufficiently regular
solutions of the flat Vlasov-Poisson system. 
As regards $\Phi$, we assume for the moment that 
\[
\Phi\in C^{1}([0,\infty[)\ \mbox{is strictly convex},\ \Phi(0)=\Phi'(0)=0,\
\lim\limits_{\eta\rightarrow\infty}\Phi(\eta)/\eta=\infty.
\] 
These assumptions make $\Phi$ non-negative and $\Phi'$ a bijection on 
$[0,\infty[$.

Our aim is to show that the energy-Casimir functional 
$\mathcal{H}_{\mathcal{C}}$ has a minimizer in the constraint set
\[
\mathcal{F}_{M}:=\left\{f\in L_{+}^{1}(\mathbb{R}^{4}) \mid
E_{\mathrm{kin}}(f) + \mathcal{C}(f)<\infty,\ \rho_{f}\in L^{4/3}(\mathbb{R}^{2}), 
\iint f=M\right\},
\]
where $M>0$ is prescribed, and the subscript $+$ indicates that only
non-negative functions are considered. 
Since the troublesome term in the functional
is the potential energy which actually depends only on the spatial
density induced by $f$ 
we introduce a reduced variational problem
for a functional which is defined in terms of 
spatial densities $\rho$. 
For $r\ge  0$ we define
\[
\mathcal{G}_{r}:=
\left\{g\in L^{1}_{+}(\mathbb{R}^{2}) \mid
\int\left(\frac{1}{2}|v|^{2}g(v)+\Phi(g(v))\right)\,\mathrm{d}v<\infty,
\int g(v)\,\mathrm{d}v=r\right\}
\]
and
\[
\Psi(r):=\inf\limits_{g\in\mathcal{G}_{r}}\int\left(\frac{1}{2}|v|^{2}g(v)+
\Phi(g(v))\right)\,\mathrm{d}v.
\]
The idea behind this construction is to first minimize the energy-Casimir
functional over all functions $f(x,v)$ which upon integration in $v$
give the same spatial density $\rho$, and then minimize with respect to the
latter in a second (and main) step. This approach was
introduced in \cite{reinred,Wo}. 

The reduced variational problem is to minimize the reduced functional
\[
\mathcal{H}_{\mathcal{C}}^{\mathrm{r}}(\rho):=\int\Psi(\rho(x))\,\mathrm{d}x+
E_{\mathrm{pot}}(\rho)
\]
over the set
\[
\mathcal{F}_{M}^{\mathrm{r}}
:=\left\{\rho\in L^{4/3}\cap L_{+}^{1}(\mathbb{R}^{2}) \mid
\int\Psi(\rho(x))\,\mathrm{d}x<\infty,\int\rho(x)\,\mathrm{d}x=M\right\}.
\]
We need to establish a relation between minimizers of the original
functional and minimizers of the reduced one. Here we can essentially follow
the corresponding results proven for the three dimensional case in \cite{reinred}.
First of all we explore the relation between $\Phi$ and $\Psi$.  
For a function $h:\mathbb{R}\to]-\infty,\infty]$ we denote by
\[
h^{\ast}(\lambda):=\sup_{r\in\mathbb{R}}\,(\lambda r-h(r))
\]
its Legendre transform. In what follows constants denoted by $C$ are
always positive, may depend on $\Phi$ and $M$, and may change their value
from line to line.
\begin{lem} \label{PhiPsi}
Let $\Phi$ and $\Psi$ be as specified respectively defined above,
and extend both functions by $+\infty$ to the interval
$]-\infty,0]$. Then the following holds:
\begin{itemize}
\item[\rm{(a)}] 
For $\lambda\in\mathbb{R}$,
\[
\Psi^{\ast}(\lambda)
=\int \Phi^{\ast}\left(\lambda-\frac{1}{2}|v|^{2}\right)\,\mathrm{d}v,
\]
and in particular $\Phi^{\ast}(\lambda)=0=\Psi^{\ast}(\lambda)$ for all $\lambda<0$.
\item[\rm{(b)}] 
$\Psi\in C^{1}([0,\infty[)$ is strictly convex, and $\Psi(0)=\Psi'(0)=0$.
\item[\rm{(c)}] 
Let $k>0$ and $n=k+1$.
\begin{itemize}
\item[\rm{(i)}] 
If $\Phi(f)=C\,f^{1+1/k}$ for $f\ge 0$, 
then $\Psi(\rho)=C\rho^{1+1/n}$ for  $\rho\ge 0$.
\item[\rm{(ii)}] 
If $\Phi(f)\ge Cf^{1+1/k}$ for $f\ge 0$ large, 
then $\Psi(\rho)\ge C\rho^{1+1/n}$ for $\rho\ge 0$ large.
\item[\rm{(iii)}] 
If $\Phi(f)\le Cf^{1+1/k}$ for $f\ge 0$ small, 
then $\Psi(\rho)\le C\rho^{1+1/n}$ for $\rho\ge 0$ small.
\end{itemize}
If the restriction to large or small values of $f$ can be dropped then 
the corresponding restriction for $\rho$ can be dropped as well.
\end{itemize}
\end{lem}
\prf
By definition
\begin{align*}
\Psi^{\ast}(\lambda)
&=\sup_{r\ge 0}\left[\lambda r-\inf_{r\in\mathcal{G}_r}
\int\left(\frac{1}{2}|v|^{2}g(v)+\Phi(g(v))\right)\,\mathrm{d}v\right]\\
&=\sup_{r\ge 0}\sup_{g\in\mathcal{G}_{r}}
\int\left[\left(\lambda-\frac{1}{2}|v|^{2}\right)g(v)-\Phi(g(v))\right]
\,\mathrm{d}v\\
&=\sup_{g\in L^{1}_{+}(\mathbb{R}^{2})}
\int\left[\left(\lambda-\frac{1}{2}|v|^{2}\right)g(v)-\Phi(g(v))\right]
\,\mathrm{d}v\\
&\leq \int\sup_{y\ge 0}\left[\left(\lambda-\frac{1}{2}|v|^{2}\right)y-\Phi(y)\right]
\,\mathrm{d}v
=\int \Phi^{\ast}\left(\lambda-\frac{1}{2}|v|^{2}\right)\,\mathrm{d}v.
\end{align*}
For $\lambda\le 0$ both sides of this estimate are zero, 
so consider $\lambda >0$.
If $|v|\ge\sqrt{2\lambda}$ then $\sup_{y\ge 0}[\cdots]=0$ and for 
$|v|<\sqrt{2\lambda}$ the supremum is attained at 
$y=y_{v}:=(\Phi')^{-1}(\lambda-\frac{1}{2}|v|^{2})$.
Hence with the definition
\[
g_{0}(v):=\left\{
\begin{array}{lcr}
y_{v}&\mathrm{for}&|v|<\sqrt{2\lambda}\\
0&\mathrm{for}&|v|\ge\sqrt{2\lambda}
\end{array}
\right.,
\]
we obtain the reversed estimate, and part (a) is established.
Part~(b) is standard for Legendre transforms,
and we refer to \cite[Lemma~2.2]{reinred} for the details.
As to (c), if we assume that
$\Phi(f)\ge Cf^{1+1/k}$ for $f\ge 0$ large, we find that
$\Phi(f)\ge Cf^{1+1/k}-C'$ for $f\ge 0$. Hence for $\lambda\ge 0$,
\[
\Phi^{\ast}(\lambda)=\sup_{f\ge 0}\left(\lambda f-\Phi(f)\right)
\le C'+\sup_{f\ge 0}\left(\lambda f-C f^{1+\frac{1}{k}}\right)\leq
C + C\,\lambda^{k+1},
\]
and
\begin{align*}
\Psi^{\ast}(\lambda)
&=\int_{|v|\le\sqrt{2\lambda}}
\Phi^{\ast}\left(\lambda-\frac{1}{2}|v|^{2}\right)\,\mathrm{d}v
\le
C\,\int_{|v|\le\sqrt{2\lambda}}
\left[1+\left(\lambda-\frac{1}{2}|v|^{2}\right)^{k+1}\right]\,\mathrm{d}v\\
&\leq C\lambda+C\int_{|v|\le\sqrt{2\lambda}}
\left(\lambda-\frac{1}{2}|v|^{2}\right)^{k+1}\,\mathrm{d}v
\leq
C+C\lambda^{k+2}=C+C\lambda^{1+n}.
\end{align*}
Using the fact that $\Psi^{\ast\ast}=\Psi$ we obtain the estimate
\[
\Psi(\rho)=\sup_{\lambda\ge 0}(\rho\lambda-\Psi^{\ast}(\lambda))\ge -C'
+\sup_{\lambda\ge 0}(\rho\lambda-C\lambda^{1+n})=C\rho^{1+1/n}-C',
\]
which proves (c) (ii). The remaining estimates are shown in a similar way.
\prfe
The relation between the minimizers 
of $\mathcal{H}_{\mathcal{C}}$ and $\mathcal{H}_{\mathcal{C}}^{\mathrm{r}}$
is as follows.
\begin{thm} \label{lifting}
\begin{itemize}
\item[\rm{(a)}] 
For every function $f\in\mathcal{F}_{M}$,
\[
\mathcal{H}_{\mathcal{C}}(f)\geq \mathcal{H}_{\mathcal{C}}^{\mathrm{r}}(\rho_{f}),
\]
with equality if $f$ is a minimizer of $\mathcal{H}_{\mathcal{C}}$ 
over $\mathcal{F}_{M}$.
\item[\rm{(b)}] 
Let $\rho_{0}\in\mathcal{F}_{M}^{\mathrm{r}}$ be a minimizer of 
$\mathcal{H}_{\mathcal{C}}^{\mathrm{r}}$ with induced potential $U_{0}$. 
Then there exists a Lagrange multiplier $E_{0}\in\mathbb{R}$ such 
that the identity 
\[
\label{lagrange}
\rho_{0}=\left\{\begin{array}{ccr}
(\Psi')^{-1}(E_{0}-U_{0})&,&U_{0}<E_{0}\\
0&,&U_{0}\ge E_{0}
\end{array}
\right.
\]
holds almost everywhere. The function
\[
f_{0}:=\left\{\begin{array}{ccr}
(\Phi')^{-1}(E_{0}-E)&,&E<E_{0}\\
0&,&E\ge E_{0}
\end{array}
\right.
\ \mbox{with}\  E=E(x,v):=\frac{1}{2}\,|v|^{2} +U_{0}(x)
\]
is a minimizer of $\mathcal{H}_{\mathcal{C}}$ in $\mathcal{F}_{M}$.
\end{itemize}
\end{thm}
\prf
Since the proof follows the same lines as \cite[Thm 2.1]{reinred}
we only indicate the main arguments. The estimate in part (a)
follows directly from the definitions. Next one can show that
if $f\in \mathcal{F}_{M}$ is such that up to sets of measure zero,
\begin{equation} \label{euleq}
\Phi'(f) = E_0-E > 0\ \mbox{where}\ f>0 ,\ \mbox{and}\
E_0-E \leq 0\ \mbox{where}\ f = 0 .
\end{equation}
with $E$ defined as in (b) but with $U_f$ instead of $U_0$ 
and $E_0$ a constant, then equality holds in part (a).
If $f$ is a minimizer of $\mathcal{H}_{\mathcal{C}}$,
then the Euler-Lagrange equation implies that $f$ is of the above
form for some Lagrange multiplier $E_0$, and equality holds in (a).
The relation of $\rho_0$ and $U_0$ in part (b) is nothing but
the Euler-Lagrange equation for the reduced variational problem.
If $f_0$ is defined as in (b) then $\rho_0 = \rho_{f_0}$, in particular,
$f_0 \in \mathcal{F}_{M}$, and (\ref{euleq}) holds by definition of $f_0$.
Hence equality holds in (a) for $f_0$ so that by part (a) for any other 
$f \in \mathcal{F}_{M}$,
\[
\mathcal{H}_{\mathcal{C}}(f) \geq \mathcal{H}_{\mathcal{C}}^{\mathrm{r}}(\rho_f) 
\geq \mathcal{H}_{\mathcal{C}}^{\mathrm{r}} (\rho_0) 
= \mathcal{H}_{\mathcal{C}} (f_0),
\]
which means that $f_0$ minimizes $\mathcal{H}_{\mathcal{C}}$.  
\prfe

\smallskip

\noindent
{\bf Remark.}
(a) 
In the next section we show that under suitable assumptions on
$\Psi$ which can be translated into corresponding assumptions
on $\Phi$ the reduced variational problem has a solution
$\rho_0$. The minimizer $f_0$ obtained by the lifting
procedure in part (b) of the theorem depends only on the particle energy $E$.
The latter is for the time-independent potential $U_0$ constant
along characteristics of the Vlasov equation,
and hence $f_0$ is a steady state of the flat Vlasov-Poisson system.

\smallskip

\noindent
(b) 
If $\mathcal{H}_{\mathcal{C}}^{\mathrm{r}}$ has at least one minimizer 
in $\mathcal{F}_{M}^{\mathrm{r}}$ and if $f_{0}\in\mathcal{F}_{M}$ 
is a minimizer of 
$\mathcal{H}_{\mathcal{C}}$, then one can show that
$\rho_{0}:=\rho_{f_{0}}\in\mathcal{F}_{M}^{\mathrm{r}}$ 
is a minimizer of $\mathcal{H}_{\mathcal{C}}^{\mathrm{r}}$. This map is one-to-one
between the sets of minimizers of $\mathcal{H}_{\mathcal{C}}$ in
$\mathcal{F}_{M}$ and $\mathcal{H}_{\mathcal{C}}^{\mathrm{r}}$ 
in $\mathcal{F}_{M}^{\mathrm{r}}$ 
and is inverse to the mapping $\rho_{0}\mapsto f_{0}$ described in part (b)
of the theorem.
\section{Existence of a solution to the reduced variational problem}
In the present section we prove that the reduced energy-Casimir
functional $\mathcal{H}_{\mathcal{C}}^{\mathrm{r}}$ has a minimizer 
in the constraint set
\[
\mathcal{F}_{M}^{\mathrm{r}}:=\left\{\rho\in L_{+}^{1}(\mathbb{R}^{2})
\mid \int\Psi(\rho(x))\,\mathrm{d}x<\infty,\int\rho(x)\,\mathrm{d}x=M\right\},
\]
where $M>0$ is prescribed and $\Psi$ satisfies the assumptions
$\Psi\in C^{1}([0,\infty[)$, $\Psi(0)=\Psi'(0)=0$ and
\begin{itemize}
\renewcommand{\theenumi}{$\Psi$\arabic{enumi}}
\item[$(\Psi 1)$]
 $\Psi$ is strictly convex,
\item[$(\Psi 2)$]
 $\Psi(\rho)\ge C\rho^{1+1/n}$ for $\rho\ge 0$ large,
\item[$(\Psi 3)$]
 $\Psi(\rho)\le C\rho^{1+1/n'}$ for $\rho\ge 0$ small, 
\end{itemize}
with growth rates $n, n' \in ]0,2[$.
The core of the proof is a concentration-compactness argument to show that
along a minimizing sequence the matter cannot spread out but has to remain 
concentrated in a finite region of space. First however we show that the
energy-Casimir functional is bounded from below in such a way that
minimizing sequences are bounded in a suitable $L^p$ space.
\begin{lem}
\label{omezenost}
Under the above assumptions on $\Psi$ and for 
$\rho\in\mathcal{F}_{M}^{\mathrm{r}}$,
\[
\int \rho^{1+1/n}\,\mathrm{d}x \leq C + C \int\Psi(\rho)\,\mathrm{d}x,
\]
and
\[
\mathcal{H}_{\mathcal{C}}^{\mathrm{r}} (\rho)
\ge
\int\Psi(\rho)\,\mathrm{d}x-C-C\left(\int\Psi(\rho)\,\mathrm{d}x\right)^{n/2}.
\]
In particular,
\[
h_{M}^{\mathrm{r}}
:=\inf_{\mathcal{F}_{M}^{\mathrm{r}}}\mathcal{H}_{\mathcal{C}}^{\mathrm{r}}
>-\infty.
\]
\end{lem}
\prf
The first estimate follows by assumption $(\Psi 2)$ and the fact that
$\int \rho = M$.
By Lemma~\ref{epotes} and interpolation,
\[
-E_{\mathrm{pot}}(\rho)
\le C||\rho||_{4/3}^{2}\le C||\rho||_{1}^{(3-n)/2}||\rho||_{1+1/n}^{(n+1)/2}
\le C+C\left(\int\Psi(\rho)\,\mathrm{d}x\right)^{n/2},
\]
and since $0<n<2$ the proof is complete.
\prfe
We note an immediate corollary.
\begin{cor}
\label{omezmin}
Any minimizing sequence of $\mathcal{H}^{\mathrm{r}}_{\mathcal{C}}$ 
in $\mathcal{F}_{M}^{\mathrm{r}}$ 
is bounded in $L^{1+1/n}(\mathbb{R}^{2})$ and therefore contains a subsequence
which converges weakly in that space.
\end{cor}
The concentration-compactness argument mentioned above relies on the
behavior of $\mathcal{H}^{\mathrm{r}}_{\mathcal{C}}(\rho)$ if $\rho$ is scaled or
split into several parts. We start with the latter; in the sequel
$B_R$ denotes the open ball of radius $R>0$ about the origin.
\begin{lem}
\label{shifting}
Let $\rho\in\mathcal{F}_{M}^{\mathrm{r}}$. Then for $R>1$,
\[
\sup_{a\in\mathbb{R}^{2}}\int_{a+B_{R}}\!\!\rho(x)\,\mathrm{d}x
\ge\frac{1}{RM}\left(-2E_{\mathrm{pot}}(\rho)-M^{2} R^{-1}-
C||\rho||_{1+1/n}^{2} R^{-(3-n)/(n+1)}\right).
\]
\end{lem}
\prf
We split the potential energy as follows:
\begin{eqnarray*}
-2 E_{\mathrm{pot}}
&=&
\iint_{|x-y|\le 1/R}\frac{\rho(x)\rho(y)}{|x-y|}\,\mathrm{d}x\,\mathrm{d}y
+\iint_{1/R<|x-y|< R}\cdots+\iint_{|x-y|\ge R}\cdots\\
&=:&
I_{1}+I_{2}+I_{3}.
\end{eqnarray*}
By H\" older's and Young's inequalities we obtain estimates
\begin{align*}
I_{1}
&\le||\rho||_{1+1/n}||\rho\ast(\mathbf{1}_{B_{1/R}}1/|\cdot|)||_{n+1}
\le||\rho||_{1+1/n}^{2}||\mathbf{1}_{B_{1/R}}1/|\cdot|||_{(n+1)/2}\\
&\le C||\rho||_{1+1/n}^{2}R^{-(3-n)/(n+1)},\\
I_{2}&\le R\iint_{|x-y|\le R}\rho(x)\rho(y)\,\mathrm{d}x\,\mathrm{d}y=
MR\sup_{a\in\mathbb{R}^{2}}\int_{a+B_{R}}\rho(x)\,\mathrm{d}x,\\
I_{3}&\le M^{2}R^{-1}.
\end{align*}
We insert these estimates into the formula for $-2 E_{\mathrm{pot}}$ and
rearrange terms to obtain the assertion.
\prfe
Next we investigate the behavior of the reduced functional under scalings.
\begin{lem}
\label{scaling}
\begin{itemize}
\item[\rm{(a)}] 
For every $M>0$,  $h_{M}^{\mathrm{r}}<0$.
\item[\rm{(a)}] 
For every $0<\overline{M} \leq M$ the estimate
$h_{\overline{M}}^{\mathrm{r}}\ge(\overline{M}/M)^{3/2}h_{M}^{\mathrm{r}}$ holds.
\end{itemize}
\end{lem}
\prf
For $\rho\in\mathcal{F}_{M}^{\mathrm{r}}$ and $a,b>0$ we define
$\bar\rho(x):=a\rho(bx)$. 
Then
\begin{eqnarray*}
\int\bar\rho\,\mathrm{d}x
&=&
ab^{-2}\int\rho\,\mathrm{d}x,\\
E_{\mathrm{pot}}(\bar\rho)
&=&
a^{2}b^{-3}E_{\mathrm{pot}}(\rho),\\
\int\Psi(\bar\rho)\,\mathrm{d}x
&=&
b^{-2}\int\Psi(a \rho)\,\mathrm{d}x.
\end{eqnarray*}
To prove part (a) we fix a bounded and compactly supported function 
$\rho\in\mathcal{F}_{M}^{\mathrm{r}}$ and choose
$a=b^{2}$ so that $\bar\rho\in\mathcal{F}_{M}^{\mathrm{r}}$ as well. 
By $(\Psi 3)$ and since $2/n'>1$,
\[
\mathcal{H}_{\mathcal{C}}^{\mathrm{r}}(\bar\rho)=
b^{-2}\int\Psi(b^{2}\rho)\,\mathrm{d}x+bE_{\mathrm{pot}}(\rho) 
\le
Cb^{2/n'}+bE_{\mathrm{pot}}(\rho)<0
\]
for $b$ sufficiently small, and part (a) is established. 
As to part (b), we take $a=1$ and
$b=(M/\overline{M})^{1/2}\ge 1$. For this choice of parameters the mapping
$\mathcal{F}_{M}^{\mathrm{r}} \ni \rho \mapsto \bar\rho
\in\mathcal{F}_{\overline{M}}^{\mathrm{r}}$ 
is one-to-one and onto, and the estimate
\begin{align*}
\mathcal{H}_{\mathcal{C}}^{\mathrm{r}}(\bar\rho)
&=b^{-2}\int\Psi(\rho)\,\mathrm{d}x+b^{-3}E_{\mathrm{pot}}(\rho)\\
&\ge b^{-3}\left(\int\Psi(\rho)\,\mathrm{d}x
+E_{\mathrm{pot}}(\rho)\right)=
\left(\frac{\overline{M}}{M}\right)^{3/2}
\mathcal{H}_{\mathcal{C}}^{\mathrm{r}}(\rho)
\end{align*}
proves the assertion of part (b).
\prfe
\begin{cor}
\label{corshift}
Let $(\rho_{i})\subset\mathcal{F}_{M}^{\mathrm{r}}$ be a minimizing sequence of 
$\mathcal{H}_{\mathcal{C}}^{\mathrm{r}}$. 
Then there exist $\delta_{0}>0$, $R_{0}>0$, 
and a sequence of shift vectors $(a_{i})\subset\mathbb{R}^{2}$ such that
for $i$ sufficiently large,
\[
\int_{a_{i}+B_{R_0}}\rho_{i}(x)\,\mathrm{d}x\ge\delta_{0}.
\]
\end{cor}
\prf
By Corollary~\ref{omezmin}, $(||\rho_{i}||_{1+1/n})$ is bounded. 
By Lemma~\ref{scaling}~(a),
\[
E_{\mathrm{pot}}(\rho_{i})
\le
\mathcal{H}_{\mathcal{C}}^{\mathrm{r}}(\rho_{i})
\le
\frac{1}{2}h_{M}^{\mathrm{r}}<0
\]
for $i$ sufficiently large, and the assertion follows by Lemma~\ref{shifting}.
\prfe
This corollary only shows that along a minimizing sequence not all
matter can spread uniformly. In the proof of the existence theorem below
we shall actually see that the matter remains within a ball of finite
radius up to spatial shifts and an arbitrarily small remainder.
In such a situation we have the following compactness result:
\begin{lem}
\label{conclem}
Let $(\rho_{i}) \subset \mathcal{F}_{M}^{\mathrm{r}}$ be such that
\[
\rho_{i}\rightharpoonup\rho_{0}\ {\rm weakly}\ {\rm in}\
L^{1+1/n}(\mathbb{R}^{2})
\]
and such that the following concentration property holds: 
\[
\forall \epsilon > 0\; \exists R>0:\ 
\limsup_{i\to\infty}\int_{|x|> R}\rho_{i}(x)\,\mathrm{d}x < \epsilon.
\]
Then
\[
E_{\mathrm{pot}}(\rho_{i}-\rho_{0})\rightarrow 0\ 
{\rm and}\ E_{\mathrm{pot}}(\rho_{i})\rightarrow E_{\mathrm{pot}}(\rho_{0}),\
i\rightarrow\infty.
\]
\end{lem}
\prf
By weak convergence $\rho_0\geq 0$ a.~e., and $\int \rho_0 \leq M$.
We define $\sigma_{i}:=\rho_{i}-\rho_{0}$ so that $\sigma_i \rightharpoonup 0$
weakly in $L^{1+1/n}(\mathbb{R}^{2})$, the concentration property holds for
$|\sigma_i|$ as well, and $\int |\sigma_i| \leq 2 M$. We need to prove that
\[
I_{i}:=\iint\frac{\sigma_{i}(x)\sigma_{i}(y)}{|x-y|}\,\mathrm{d}x\,\mathrm{d}y\to
0,
\]
which is the first assertion. Since
\[
E_{\mathrm{pot}}(\rho_i) - E_{\mathrm{pot}}(\rho_0)
= E_{\mathrm{pot}}(\rho_i-\rho_0) + \int U_{\rho_0} (\rho_i - \rho_0),
\]
the fact that $U_{\rho_0} \in L^4(\mathbb{R}^2)$ together with the 
weak convergence of $\rho_i$ implies the second assertion.
For $\delta >0$ and $R>0$ we split the domain of integration 
into three subsets defined by
\begin{eqnarray*}
&&|x-y|<\delta,\\
&&|x-y|\ge \delta  \wedge (|x|\ge R \vee |y|\ge R),\\
&&|x-y|\ge \delta  \wedge |x|<R \wedge |y|<R,
\end{eqnarray*}
and we denote the corresponding contributions to 
$I_{i}$ by $I_{i,1}$, $I_{i,2}$, $I_{i,3}$.
Young's inequality implies that
\[
|I_{i,1}|\le
C||\sigma_{i}||_{1+1/n}^{2}||\mathbf{1}_{B_{\delta}}|\cdot|^{-1}||_{(n+1)/2}
\le C\left(\int_{0}^{\delta} r^{(1-n)/2}\,\mathrm{d}r\right)^{2/(n+1)}
\]
which can be made as small as we wish, uniformly in $i$ and
$R>0$, by making $\delta >0$ small.  For $\delta>0$ now fixed,
\[
|I_{i,2}|\le\frac{4 M}{\delta} \int_{|x|>R}|\sigma_{i}(x)|\,\mathrm{d}x
\]
which becomes small for $i \to \infty$
by the concentration assumption, if we choose $R>0$ accordingly.
Finally by H\" older's inequality,
\[
|I_{i,3}|=\left|\int\sigma_{i}(x)h_{i}(x)\,\mathrm{d}x\right|
\le ||\sigma_{i}||_{1+1/n}||h_{i}||_{1+n}\le C||h_{i}||_{1+n},
\]
where in a pointwise sense,
\[
h_{i}(x):=\mathbf{1}_{B_{R}}(x)\int_{|x-y|\ge\delta}\mathbf{1}_{B_{R}}(y)
\frac{1}{|x-y|}\sigma_{i}(y)\,\mathrm{d}y \to 0
\]
due to the weak convergence of $\sigma_i$ and the fact that the test function
against which $\sigma_i$ is integrated here is in $L^{1+n}$.
Since $|h_i| \leq \frac{2 M}{\delta} \mathbf{1}_{B_R}$ uniformly in $i$,
Lebesgue's dominated convergence theorem implies that $h_i \to 0$
in $L^{1+n}$, and the proof is complete.
\prfe
We have now assembled all the tools we need to prove the existence of
a minimizer of the reduced functional.
\begin{thm}
\label{mainthm}
Let $(\rho_{i})\subset\mathcal{F}_{M}^{\mathrm{r}}$ be a minimizing sequence
of $\mathcal{H}_{\mathcal{C}}^{\mathrm{r}}$. Then there exists a sequence of 
shift vectors $(a_{i})\subset\mathbb{R}^{2}$ and a subsequence, 
again denoted by $(\rho_{i})$, such that for every $\varepsilon>0$ 
there exist $R>0$ with
\[
\int_{a_{i}+B_{R}}\rho_{i}(x)\,\mathrm{d}x \ge M-\varepsilon,\ i\in\mathbb{N},
\]
\[
T_{a_{i}}\rho_{i}:=\rho_{i}(\cdot+a_{i})\rightharpoonup\rho_{0}\ 
{\rm weakly}\ {\rm in}\ L^{1+1/n}(\mathbb{R}^{2}),\ i\to\infty,
\]
\[
\int_{B_{R}}\rho_{0}(x)\,\mathrm{d}x \ge M-\varepsilon.
\]
Finally,
\[
E_{\mathrm{pot}}(T_{a_{i}}\rho_{i}-\rho_{0})\to 0,
\]
and $\rho_{0}\in\mathcal{F}_{M}^{\mathrm{r}}$ is a minimizer of 
$\mathcal{H}_{\mathcal{C}}^{\mathrm{r}}$.
\end{thm}
\prf
We split $\rho\in\mathcal{F}_{M}^{\mathrm{r}}$ as follows:
\[
\rho=\mathbf{1}_{B_{R_{1}}}\rho+\mathbf{1}_{B_{R_{2}}\setminus B_{R_{1}}}\rho+
\mathbf{1}_{\mathbb{R}^2\setminus B_{R_{2}}}\rho=:\rho_{1}+\rho_{2}+\rho_{3}.
\]
The parameters $R_{1}<R_{2}$ of the split are yet to be determined. 
Recalling the definition of the bilinear form 
$\langle\cdot,\cdot\rangle_\mathrm{pot}$,
\[
\mathcal{H}_{\mathcal{C}}^{\mathrm{r}}(\rho)
=\mathcal{H}_{\mathcal{C}}^{\mathrm{r}}(\rho_{1})
+\mathcal{H}_{\mathcal{C}}^{\mathrm{r}}(\rho_{2})
+\mathcal{H}_{\mathcal{C}}^{\mathrm{r}}(\rho_{3})
- 2 \langle \rho_1+\rho_3,\rho_2\rangle_\mathrm{pot}
- 2 \langle \rho_1,\rho_3\rangle_\mathrm{pot}.
\]
If we choose $R_{2}>2R_{1}$, then
\[
\langle \rho_1,\rho_3\rangle_\mathrm{pot} \le\frac{C}{R_{2}}.
\]
By Lemma~\ref{epotes} and interpolation,
\begin{eqnarray*}
\langle \rho_1+\rho_3,\rho_2\rangle_\mathrm{pot}
&\leq&
||\rho_1+\rho_3||_\mathrm{pot}
||\rho_2||_\mathrm{pot}\\
&\le& 
C\, ||\rho_{1}+\rho_{3}||_{4/3}||\rho_{2}||_{\mathrm{pot}}
\leq
C\,||\rho||_{1+1/n}^{(n+1)/4}||\rho_{2}||_{\mathrm{pot}}.
\end{eqnarray*}
If we define
\[
M_{l}:=\int\rho_{l}(x)\,\mathrm{d}x,\ l=1,2,3,
\]
then Lemma~\ref{scaling}~(b) and the estimates above imply that
\begin{eqnarray*}
h_{M}^{\mathrm{r}}-\mathcal{H}_{\mathcal{C}}^{\mathrm{r}}(\rho)
&\le&
\left(1-\left(\frac{M_{1}}{M}\right)^{3/2}-\left(\frac{M_{2}}{M}\right)^{3/2}
-\left(\frac{M_{3}}{M}\right)^{3/2}\right)h_{M}^{\mathrm{r}}\\
&&
{}+C\left(R_{2}^{-1}+||\rho||_{1+1/n}^{(n+1)/4}
||\rho_{2}||_{\mathrm{pot}}\right)\\
&\le&
\frac{C}{M^{2}}(M_{1}M_{2}+M_{2}M_{3}+M_{1}M_{3})h_{M}^{\mathrm{r}}\\
&&
{}+C\left(R_{2}^{-1}+||\rho||_{1+1/n}^{(n+1)/4}
||\rho_{2}||_{\mathrm{pot}}\right)\\
&\le& 
C h_{M}^{\mathrm{r}}M_{1}M_{3}+C\left(R_{2}^{-1}+
||\rho||_{1+1/n}^{(n+1)/4}||\rho_{2}||_{\mathrm{pot}}\right).
\end{eqnarray*}
Here we used that for some constant $C>0$ the following inequality holds:
\[
x^{3/2} + y^{3/2} + z^{3/2} \leq 1 - C (x y + x z + y z)\ \mbox{for}\ x, y, z
\geq 0
\ \mbox{with}\ x + y + z = 1. 
\]
Now we consider a minimizing sequence 
$(\rho_i) \subset\mathcal{F}_{M}^{\mathrm{r}}$ of 
$\mathcal{H}_{\mathcal{C}}^{\mathrm{r}}$ and choose shift vectors 
$(a_i)\subset \mathbb{R}^{2}$, $\delta_0>0$, and $R_0>0$ according to
Cor.~\ref{corshift}. 
Since all our functionals are invariant under spatial translations
the sequence $T_{a_{i}}\rho_{i}=\rho_{i}(\cdot+a_{i})$ is again minimizing
and hence bounded in $L^{1+1/n}(\mathbb{R}^{2})$ so that up to a subsequence
we can assume that it converges weakly to some 
$\rho_{0} \in L^{1+1/n}(\mathbb{R}^{2})$. 
We choose $R_{1}>R_{0}$ so that by Cor.~\ref{corshift}, $M_{i,1}\ge\delta_{0}$
for $i$ large, and
\[
-Ch_{M}^{\mathrm{r}}\delta_{0}M_{i,3}
\le
C\,R_{2}^{-1}+C||\rho_{0,2}||_{\mathrm{pot}}
+C||\rho_{i,2}-\rho_{0,2}||_{\mathrm{pot}}+
\mathcal{H}_{\mathcal{C}}^{\mathrm{r}}(T_{a_{i}}\rho_{i})-h_{M}^{\mathrm{r}}.
\]
Given any $\varepsilon>0$ we increase $R_{1}>R_{0}$ such that the second term
on the right hand side is smaller than $\varepsilon$. 
Next we choose $R_{2}>2R_{1}$ such that the first term is small. 
Now that $R_{1}$ and $R_{2}$ are fixed, the third term  converges 
to zero by Lemma \ref{conclem}, and since $T_{a_{i}}\rho_{i}$ is minimizing 
the remainder follows suit. Therefore for $i$ sufficiently large,
\[
\int_{B_{R_{2}}}T_{a_{i}}\rho_{i}\,\mathrm{d}x
= M-M_{i,3}
\ge M-(-Ch_{M}^{\mathrm{r}}\delta_{0})^{-1}\varepsilon.
\]
The strong convergence of the potential energies now follows by
Lemma \ref{conclem}.
By weak convergence  $\rho_{0}\ge 0$ a.e., and for any $\varepsilon >0$ there
exists $R>0$ such that
\[
M\ge\int_{B_{R}}\rho_{0}\,\mathrm{d}x\ge M-\varepsilon,
\]
in particular $\rho_{0}\in L^{1}(\mathbb{R}^{2})$ with $\int\rho_{0}=M$. 
The functional $\rho\mapsto\int\Psi(\rho)\,\mathrm{d}x$ is convex, 
so by Mazur's lemma \cite[2.13]{LL} and Fatou's lemma
\[
\int\Psi(\rho_0)\,\mathrm{d}x
\le\limsup_{i\to\infty}\int\Psi(T_{a_{i}}\rho_{i})\,\mathrm{d}x.
\]
Hence $\rho_{0} \in \mathcal{F}_{M}^{\mathrm{r}}$ with
\[
\mathcal{H}_{\mathcal{C}}^{\mathrm{r}}(\rho_{0})\le
\limsup_{i\to\infty}\mathcal{H}_{\mathcal{C}}^{\mathrm{r}}(\rho_{i})
=h_{M}^{\mathrm{r}},
\]
and the proof is complete.
\prfe

\smallskip

\noindent
{\bf Remark.} 
(a)
Thm.~\ref{mainthm} provides a minimizer $\rho_0$
of the reduced energy-Casimir functional 
$\mathcal{H}_{\mathcal{C}}^{\mathrm{r}}$
under the assumptions
$(\Psi 1)$--$(\Psi 3)$. By Thm.~\ref{lifting} this minimizer can be lifted to
a minimizer $f_0$ of the original energy-Casimir functional
$\mathcal{H}_{\mathcal{C}}$.
By Lemma~\ref{PhiPsi}
the function $\Psi$ satisfies the necessary assumptions if
$\Phi$ which appears in the original Casimir functional
satisfies the following ones:
$\Phi\in C^{1}([0,\infty[)$, $\Phi(0)=\Phi'(0)=0$ and
\begin{itemize}
\renewcommand{\theenumi}{$\Phi$\arabic{enumi}}
\item[$(\Phi 1)$]
 $\Phi$ is strictly convex,
\item[$(\Phi 2)$]
 $\Phi(f)\ge C f^{1+1/k}$ for $f\ge 0$ large,
\item[$(\Phi 3)$]
 $\Phi(f)\le C f^{1+1/k'}$ for $f\ge 0$ small, 
\end{itemize}
with growth rates $k, k' \in ]0,1[$.

\smallskip

\noindent
(b)
As will be seen in the next section the mere fact that $f_0$
minimizes $\mathcal{H}_{\mathcal{C}}$ is not sufficient for stability.
However, let $(f_i)\subset \mathcal{F}_{M}$ be a minimizing
sequence of $\mathcal{H}_{\mathcal{C}}$. By Thm.~\ref{lifting}~(a)
the sequence of induced spatial densities $\rho_i=\rho_{f_i}$
is minimizing for $\mathcal{H}_{\mathcal{C}}^{\mathrm{r}}$.
Choose a subsequence of $(\rho_i)$ (and $(f_i)$) and
shift vectors such that the assertions of Thm.~\ref{mainthm}
hold, and denote the shifted subsequence again by $(f_i)$.
We claim that this sequence converges weakly to $f_0$.
Clearly, $(f_i)$ is bounded in $L^{1+1/k}(\mathbb{R}^{4})$
with bounded kinetic energy, and 
$E_{\mathrm{pot}}(f_i) = E_{\mathrm{pot}}(\rho_i) \to
E_{\mathrm{pot}}(\rho_0)$.
Any subsequence of $(f_i)$ must therefore have a weakly convergent
subsequence with weak limit $\tilde f_0$ which is a minimizer
of $\mathcal{H}_{\mathcal{C}}$ and induces the same spatial
density $\rho_0$ and potential $U_0$. But then by
Thm.~\ref{lifting}, $\tilde f_0 = f_0$ so that indeed
$f_i \rightharpoonup f_0$ weakly in $L^{1+1/k}(\mathbb{R}^{4})$. 

\smallskip

\noindent
(c)
For $k\geq1$ one can still obtain stability results, 
cf.~\cite{F} for the Kuzmin
disk which corresponds to $\Phi(f)=f^{3/2}$, i.e., $k=k'=2$.
However, the reduction approach cannot work, because as we shall see in
the last section this approach implies stability for the Euler-Poisson 
system where stability is probably lost at $n=2$, i.e., $k=1$.
\section{Stability of minimizers}
Now that the existence of a minimizer is proven, 
we can explore its dynamical stability properties. 
So let $\rho_0$ be as obtained in Thm.~\ref{mainthm} and
$f_0$ as induced by Thm.~\ref{lifting}. A simple expansion shows that
\begin{eqnarray}
\label{hcdif}
\mathcal{H}_{\mathcal{C}}(f)-\mathcal{H}_{\mathcal{C}}(f_{0})
=d(f,f_{0})+E_{\mathrm{pot}}(f-f_{0}),
\end{eqnarray}
where for $f\in\mathcal{F}_{M}$ and with the Lagrange multiplier $E_0$ from
Thm.~\ref{lifting}~(b),
\begin{eqnarray*}
d(f,f_{0})
&:=&
\iint[\Phi(f)-\Phi(f_{0})+E(f-f_{0})]\,\mathrm{d}v\,\mathrm{d}x\\
&=&
\iint[\Phi(f)-\Phi(f_{0})+(E-E_{0})(f-f_{0})]\,\mathrm{d}v\,\mathrm{d}x\\
&\ge &
\iint[\Phi'(f_{0})+(E-E_{0})](f-f_{0})\,\mathrm{d}v\,\mathrm{d}x\ge 0
\end{eqnarray*}
with $d(f,f_{0})=0$ iff $f=f_{0}$. 
For the positivity of $d$ we use the strict convexity of $\Phi$ and the form
of $f_{0}$ according to Thm.~\ref{lifting}~(b); by that theorem the term in
brackets vanishes on the support of $f_0$. 
We recall that 
$-E_{\mathrm{pot}}(f) = \langle \rho_f,\rho_f\rangle_{\mathrm{pot}}
=||\rho_f||_{\mathrm{pot}}^2$
defines a norm on $\rho_f$, cf.\ Lemma~\ref{epotes};
note that the right hand side in Eqn.~(\ref{hcdif}) is
$d(f,f_{0})-||\rho_f-\rho_0||_{\mathrm{pot}}^2$.
We obtain the following stability result;
$C^{2}_{\mathrm{c}}(\mathbb{R}^{4})$ denotes the space of compactly 
supported $C^2$ functions on $\mathbb{R}^{4}$.
\begin{thm}
Let $f_{0}$ be a minimizer of $\mathcal{H}_{\mathcal{C}}$ on $\mathcal{F}_{M}$
obtained from  a minimizer $\rho_{0}$ of
$\mathcal{H}_{\mathcal{C}}^{\mathrm{r}}$, and 
assume that the minimizer is unique. Then for any $\varepsilon>0$ there exists 
$\delta>0$ such that for any classical solution $[0,T[\ni t\mapsto f(t)$ 
of the flat Vlasov-Poisson system with 
$f(0)\in C^{2}_{\mathrm{c}}(\mathbb{R}^{4})\cap\mathcal{F}_{M}$ and 
$||f(0)||_{1+1/k}=||f_{0}||_{1+1/k}$ the estimate
\[
d(f(0),f_{0}) + ||\rho_{f(0)}-\rho_0||_{\mathrm{pot}}^2<\delta
\]
implies that for each $t\in[0,T[$ there exists a shift 
vector $a\in\mathbb{R}^{2}$ such that
\[
||f(t)-T_{a}f_{0}||_{1+1/k}+d(f(t),T_{a}f_{0}) 
+||\rho_{f(t)}-T_{a}\rho_{0}||_{\mathrm{pot}}^2<\varepsilon.
\]
\end{thm}
\prf
Assume that the assertion were false. 
Then there exists 
$\varepsilon >0$, $t_{j}>0$, 
$f_{j}(0)\in C^{2}_{\mathrm{c}}(\mathbb{R}^{4}) \cap\mathcal{F}_{M}$ with
$||f_{j}(0)||_{1+1/k}=||f_{0}||_{1+1/k}$ such that for every $j\in\mathbb{N}$,
\begin{equation}
\label{falsch}
d(f_{j}(0),f_{0}) - E_{\mathrm{pot}}(f_{j}(0)-f_{0})<\frac{1}{j}
\end{equation}
but for any shift vector $a\in\mathbb{R}^{2}$,
\begin{equation}
\label{spor}
||f_{j}(t_{j})-T_{a}f_{0}||_{1+1/k}+d(f_{j}(t_{j}),T_{a}f_{0}) - 
E_{\mathrm{pot}}(f_{j}(t_{j})-T_{a}f_{0})\ge\varepsilon .
\end{equation}
Since $\mathcal{H}_{\mathcal{C}}$ is preserved along solutions we have 
from (\ref{hcdif}) and (\ref{falsch}) that
\[
\mathcal{H}_{\mathcal{C}}(f_{j}(t_{j}))
=\mathcal{H}_{\mathcal{C}}(f_{j}(0))\to\mathcal{H}_{\mathcal{C}}(f_{0}),
\]
i.e., $(f_{j}(t_{j}))$ is a minimizing sequence.
By Thm.~\ref{mainthm} and the remark at the end of the previous section
there is a sequence of shift vectors 
$(a_{j})\subset\mathbb{R}^{2}$ such that up to a subsequence,
\[
\lim_{j\to\infty}E_{\mathrm{pot}}(T_{a_{j}}f_{j}(t_{j})-f_{0})\to 0.
\]
By (\ref{hcdif}) this implies that $d(T_{a_{j}}f_{j}(t_{j}),f_{0})\to 0$. 
For the convergence of ${||\cdot||_{1+1/k}}$ we use the fact that 
$||f_j(t)||_{1+1/k}=||f_j (0)||_{1+1/k} = ||f_0||_{1+1/k}$ for any $t>0$. 
By the remark,
$T_{a_{j}}f_{j}(t_{j})\rightharpoonup f_{0}$ weakly in
$L^{1+1/k}(\mathbb{R}^{4})$,
and hence $T_{a_{j}}f_{j}(t_{j})\to f_{0}$ strongly in
$L^{1+1/k}(\mathbb{R}^{4})$.
But these convergence results for $T_{a_{j}}f_{j}(t_{j})$ 
contradict (\ref{spor}).
\prfe

\smallskip

\noindent
{\bf Remark.}
(a)
The uniqueness assumption on $f_0$ in the above theorem is made mostly in
order to avoid technical complications. It suffices if $f_0$ is isolated
with respect to the topology of our stability estimate. If there should be
a continuum of minimizers then the set of minimizers itself is stable;
we refer to \cite[Thm.~4]{GR2} for such a formulation of the result
in the three dimensional case. We are not aware of a case where 
there is a continuum of minimizers with fixed mass $M$.
For a closely related variational problem it has been shown that the above 
stability estimate remains valid even then \cite{Sch2}.

\smallskip

\noindent
(b)
As opposed to the three dimensional case \cite{Pf,reinbook,Sch1} there 
is no global existence and uniqueness result to the initial value problem
for the flat Vlasov-Poisson system yet. Hence our stability result is
conditional in the sense that it holds as long as a suitable solution exists.
A local existence and uniqueness result for smooth solutions with
initial data in $C^{2}_{\mathrm{c}}(\mathbb{R}^{4})$ as well
as a global existence result for weak solutions to the flat system
was established in \cite{Shev}. We could also carry out our stability analysis 
in the framework of these global weak solutions, but this would only bury
the main ideas under technicalities.

\smallskip

\noindent
(c)
By interpolation between $L^{1}$ and $L^{1+1/k}$ we obtain a stability
estimate for $||f(t)-T_a f_0||_{p}$ with $p\in ]1,1+1/k]$. If we assume 
that the initial perturbations have supports of uniformly bounded
measure we can include the case $p=1$, if we assume a uniform bound on the 
$L^\infty$ norm of the initial perturbations we can by interpolation
include all $p\in [1+1/k,\infty[$; notice that both the measure
of the support and the $L^\infty$ norm are invariant under classical
solutions of the Vlasov-Poisson system.

\smallskip

\noindent
(d)
The need for the shifts in the stability estimate arises
from the Galilei invariance of the Vlasov-Poisson system. 
If $f_0$ is a steady state then for any fixed $V\in\mathbb{R}^{2}$
the function $f_0(x-t V, v-V)$ is a
time dependent solution;
$f_0$ is simply put into a uniformly moving coordinate system.
But while the distance of this perturbation to the steady state
grows linearly in $t$, it is arbitrarily close to the steady state
at $t=0$ for $V$ small.
\section{Connection to the Euler-Poisson system}
A self-gravitating matter distribution can be described
on the microscopic, kinetic level represented by the
Vlasov-Poisson system or on the macroscopic, fluid level
represented by the Euler-Poisson system. 
The reduction technique connects
the stability problems for these two viewpoints.  
In the three dimensional
situation this connection was observed in \cite{R7}. In the flat case
the corresponding Euler-Poisson system reads
\[
\frac{\partial\rho}{\partial t}+\mathrm{div}(\rho u)=0,
\]
\[
\rho\frac{\partial u}{\partial t}+(u\cdot\nabla_{x})u
=-\nabla_{x}p-\rho\nabla_{x}U,
\]
\[
U(t,x)=-\int\frac{\rho(t,y)}{|x-y|}\,\mathrm{d}y,
\]
with the equation of state
\[
p(\rho)=\rho\Psi'(\rho)-\Psi(\rho).
\]
Here $p$ denotes the pressure of the fluid and $u$ denotes its
velocity field; the meaning of $\rho$ and $U$ is as before.
If $\rho_0$ is a minimizer of the reduced energy-Casimir
functional $\mathcal{H}_{\mathcal{C}}^{\mathrm{r}}$, then using the
Euler-Lagrange identity in Thm.~\ref{lifting}~(b) it
is easy to check that
$\rho_{0}$ and the zero velocity field $u_{0}\equiv 0$ solve the
flat Euler-Poisson system. Clearly, the state $(\rho_{0},u_{0})$ 
minimizes the energy
\[
\mathcal{H}(\rho,u)
=\frac{1}{2}\int|u|^{2}\rho\,\mathrm{d}x
+\int\Psi(\rho)\,\mathrm{d}x+E_{\mathrm{pot}}(\rho),
\]
among the states with $\int \rho = M$.
Formally, the energy is conserved along solutions of
the Euler-Poisson system. An expansion about $(\rho_{0},u_{0})$
gives
\[
\mathcal{H}(\rho,u)-\mathcal{H}(\rho_{0},u_{0})
=\frac{1}{2}\int|u|^{2}\rho\,\mathrm{d}x+d(\rho,\rho_{0})+
E_{\mathrm{pot}}(\rho-\rho_{0}),
\]
where
\[
d(\rho,\rho_{0}):=\int[\Psi(\rho)-\Psi(\rho_{0})+
(U_{0}-E_{0})(\rho-\rho_{0})]\,\mathrm{d}x\ge 0.
\]
Now the stability proof proceeds in the same way as in the Vlasov
case. We can for every $\varepsilon>0$ find a $\delta>0$ such that
for every solution $t\mapsto(\rho(t),u(t))$ of the flat
Euler-Poisson system with
$\rho (0)\in\mathcal{F}_{M}^{\mathrm{r}}$, which preserves energy
and mass, the initial estimate
\[
\frac{1}{2}\int|u(0)|^{2}\rho(0)\,\mathrm{d}x+d(\rho(0),\rho_{0})
+||\rho(0)-\rho_{0}||_{\mathrm{pot}}^2<\delta
\]
implies that as long as the solution exists and up to shifts in space,
\[
||\rho(t)-\rho_0||_{1+1/n} + 
\frac{1}{2}\int|u(t)|^{2}\rho(t)\,\mathrm{d}x+d(\rho(t),\rho_{0})
+||\rho(t)-\rho_{0}||_{\mathrm{pot}}^2<\varepsilon.
\]
Neither in the flat case nor in the three dimensional one
is there an  existence theory for global solutions of the 
Euler-Poisson system, which preserve all the necessary quantities,
so the result is conditional in this sense.

\end{document}